# An Analytical 2-D Model of Triple Metal Double Gate Graded Channel Junctionless MOSFET with Hetero-dielectric Gate Oxide Stack

**Shib Sankar Das[1], Barun Majumder[2], Ankush Ghosh[1]**

[1] School of Engineering and Applied Science, The Neotia University, India
[2] University of Tennessee, USA

## Abstract

*In this paper, a two-dimensional analytical model of a laterally graded-channel triple-metal double-gate Junctionless Field Effect Transistor with hetero dielectric gate oxide stack consisting of $SiO_2$ and $HfO_2$ is derived. The model illustrates higher drive current and better performance against hazardous SCEs and HCEs in below 30 nm regime. Parabolic approximation method is used here to construct channel potentials and electric fields by solving 2-D Poisson's equation with applicable boundary conditions. The basic central and surface potentials as well as central and surface electric fields are being illustrated, Threshold voltage, DIBL, sub-threshold swing (SS) and a compact current model have also been deduced. These parameters clearly show the benefits of proposed graded-channel triple-metal double-gate structure with hetero-dielectric gate oxide stack. The device has reliability in low power applications because of its better $I_{on}$ and $I_{off}$ control. Finally, the analytical model is validated with self-consistent numerical calculations to illustrate better performance of among available junctionless devices.*

## 1. Introduction

Rapid growth in microelectronics industry demands more and more device miniaturization and leads to downscaling of MOSFETs into nanometre scale. By continuous shrinking of the device in last few decades the gates lose its complete control over the channel and simultaneously hazardous SCEs degrade device performance. Pioneers of Nanotechnology have been investigating for novel device structure for years to reduce the adverse SCEs and to enhance the device performance in low power applications. There are several types of proposed approaches for replacement of previous technology such as Source and Drain engineering, Material engineering, Gate and channel engineering and multi-gate structures etc. Among all of these alternative approaches, Junctionless (JL) MOSFET is one of the most hopeful devices among all alternative approaches that has not only been gladly accepted but also experimentally verified. The concept related to junctionless transistor is studied in literature [1]-[5]. In junctionless transistor p-n junctions are absent in its source-channel-drain path. By virtue of its junction free nature, it can be fabricated in simpler steps providing better scalability and thermal budget of the device and boosting better immunity to SCEs [6].

At room temperature, the restricted value of subthreshold swing of a MOSFET is 60mV/decade (ITRS 2016). Lee et al. [7] demonstrates that in comparison with any conventional MOSFET, junctionless transistor has improved subthreshold slope (SS) and

Email: barunbasanta@gmail.com

drain induced barrier lowering (DIBL) and its result enhanced performance of JL devices. Ching et al. [8] introduces a quasi-two-dimensional threshold voltage model for a short channel JLDGFET. Ping Wang et al. [9] demonstrates threshold voltage and subthreshold swing of dual metal double gate junctionless MOSFET and justify its reliability. Darwin S and Arun Samuel TS et al. [10] revealed a JL DMDG stack MOSFET which depicted improved subthreshold swing and ON-OFF ratio than single oxide layer JL DMDG MOSFET and proved its superiority Lou et al. [11] addressed dual material gate junctionless transistor and showed improved results in drain current and other parameters in comparison with single gate JL transistor. Recently an analytical modelling of uniformly doped triple metal double gate JL MOSFET has been proposed by Darwin S and Arun Samuel TS et al. [12] and their model predicts remarkably low leakage current and $\frac{I_{on}}{I_{off}}$ is $10^9$, but threshold voltage model is missing in this article.

In the context of channel engineering, Chen et al. [13] proposed a laterally graded channel model which is adversely doped with high concentration at source end and low concentration at drain end. This demonstrates the improvement of analog/RF performance in case of a DGJLFET. But this is a simulation-based paper, no analytical model is derived here. V Pathak et al. [14] demonstrates a dual material gate graded channel junctionless MOSFET for analog application and gets higher values of drain current ($I_{DS}$) in comparison with uniformly doped channel device. This paper also illustrates a simulation-based project and no analytical expression is derived. However triple metal double gate graded channel junctionless MOSFET containing laterally increasing three different electron concentrations in the channel region is not found in literature.

In contrary to gate oxide engineering S Amin and R Sarin et al. [15] reported a model of charged plasma based dual metal double gate with gate stacked (high-k gate oxide + $SiO_2$) structure of junctionless transistor for improvement of the analog performance in comparison with JL transistor containing only $SiO_2$ gate oxide. However, no such model is found in literature which describes the effect of gate stacked oxide (high-k gate oxide + $SiO_2$) in triple material double gate graded channel junctionless MOSFET.

In a junctionless transistor current flows through bulk of channel region. Duarte et al. [16] demonstrates bulk current model for a junctionless double gate transistor in long channel case and also illustrate [17] a full range drain current model for JLFET. Drain current model for a dual metal double gate junctionless transistor is also illustrated by V Kumari et al. [18]. Z-M-Lin et al. [19] describe a subthreshold current model for a junctionless double gate transistor.

The objective of this paper is to improve the reliability and performances of junctionless MOSFETs, available in the literature, through this novel device, incorporating gate and channel engineering in the desired sector of low power applications. This paper demonstrates analytical expressions for both central and surface potential, electric field, threshold voltage, DIBL, roll-off, subthreshold swing (SS) and the drain current in subthreshold, linear and saturation region exclusively for triple-metal double-gate with hetero-dielectric gate oxide stack graded-channel junctionless MOSFET that are not found in literature. To justify the betterment of this novel device some illustrations, existing in literature have compared with this new one. When a laterally graded channel has incorporated with a junctionless MOSFET, Chen et al [20] predict threshold voltage ($V_{th}$)= 0.18V for $V_{DS}$=1V and off state current, $I_{off} = 1 X 10^7$ Amp. Junctionless double metal double gate MOSFET suggested by P Wang et al. [9] estimated threshold voltage roll-off, DIBL and Subthreshold Swing (SS) for $t_{Si}$= 10 nm, $t_{ox}$=2 nm and L= 20nm as 0.29V, 0.08V/V and 135 mV/decade. Moreover, recently reported junctionless triple material double gate MOSFET suggested by Darwin S and Arun Samuel et al.[12], predicts Subthreshold Swing (SS) value 100 mV/decade for $V_{DS}$=0.5V and off state current $I_{off} = 1X\ 10^{11}$Amp for $V_{DS}$=0.7V.In this present paper, the quantitative values of $V_{th} = 0.8V$, $V_{th}$ roll-off= -0.6V, DIBL = 0.6V and $I_{off} = 1X\ 10^{18}$Amp for $V_{DS}$=0.3V and, Subthreshold swing (SS)= 90 mV/decade for $V_{GS}$=0.1V are predicted. In comparison with the above mentioned junctionless MOSFETs, this newly proposed device gets exclusive advantage in low power applications with commensurable short channel immunity and reliability. This paper comprises five different sections. Section 2 contains description of novel device structure with necessary parameter specifications. Section 3 describes physics based analytical model of proposed device. Section 4 accumulates different characteristic plots to analyse the performance of the device. Final conclusions are drawn in Section 5.

## 2. Device Structure and Parameter specification

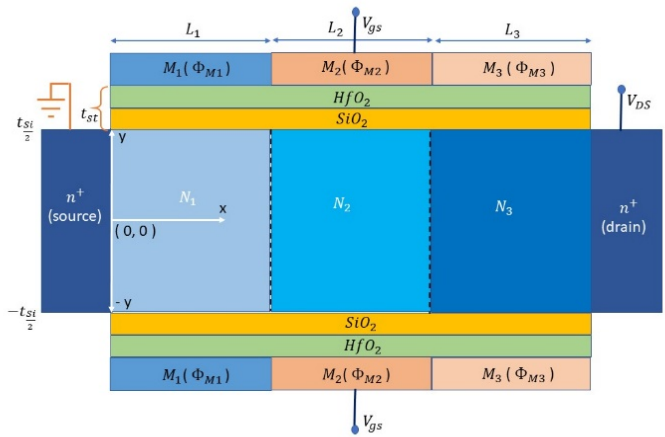

**Fig 1: Triple Metal Double Gate Graded Channel Junctionless MOSFET with Hetero-dielectric Gate Oxide stack**

Fig. 1 illustrates the schematic diagram of the proposed device structure named, triple metal double gate graded channel junctionless field effect transistor.

In this state of art, alignment of three different doping concentrations in channel region which increase laterally from source end to drain end has been done. Basic motivation of this paper is to construct a novel structure of a graded channel junctionless MOSFET which is which is solemnly ann-channel device. However, source-channel-drain regions are constructed with n-type semiconductor but there exist different doping concentrations, and the height of the doping gradients are relatively low in comparison with the Junction based MOSFET structure. The illustrated device works as a conventional depletion mode junctionless MOSFET. Incorporation of different doping concentrations in the channel region (lower concentration at the source endand higher concentration at the drain end) enhance the device characteristics like less DIBL, less leakage current etc. For simplicity, it is considered that source and drain concentrations are same as their adjacent channel region concentrations. The channel doping concentration is laterally increasing towards drain end such as $N_1$ is under gate metal $M_1$, $N_2$ is under gate metal $M_2$ and $N_3$ is under gate metal $M_3$, ($N_1 < N_2 < N_3$).

The novel structure contains triple-metal gate electrode in both sides of the channel and the gate electrodes behave like double gate. The front and back electrodes are comprised with the assembly of three different metal gates having work functions, say, $\Phi_{M1}, \Phi_{M2}$ and $\Phi_{M3}$ respectively ($\Phi_{M1} > \Phi_{M2} > \Phi_{M3}$). In the proposed structure, the highest work function gate metal i.e. $\Phi_{M1}$ is placed at the source end considering a length of $L_1$ nm and called as "control gate". The metal with intermediate work function, $\Phi_{M2}$ with a gate length $L_2$ nm is placed in the middle of the gate electrode and is known as "first screen gate". The gate metal with lowest work function value, $\Phi_{M3}$ and gate length $L_3$ nm is placed at the drain end and is known as "second screen gate". For simplicity it is assumed that, $L_1 = L_2 = L_3$ and $L_1 + L_2 + L_3 = L$.

**TABLE 1**
**SPECIFICATION OF DEVICE PARAMETERS**

| Symbol | Parameter | Values taken |
|---|---|---|
| $\Phi_{M1}$ | Gate metal ($\Phi_{M1}$) | 4.46 eV (e.g., Tungsten) |
| $\Phi_{M1}$ | Gate metal ($\Phi_{M2}$) | 4.30 eV (e.g., Zinc) |
| $\Phi_{M1}$ | Gate metal ($\Phi_{M2}$) | 4.15 eV (e.g., Aluminium) |
| $L_1$ | Gate length under $M_1$ | 9 nm |
| $L_2$ | Gate length under $M_2$ | 9 nm |
| $L_3$ | Gate length under $M_3$ | 9 nm |
| $N_1$ | Channel concentration under $M_1$ | $2 \times 10^{18}/cm^3$ |
| $N_2$ | Channel concentration under $M_2$ | $5 \times 10^{18}/cm^3$ |
| $N_3$ | Channel concentration under $M_3$ | $1 \times 10^{19}/cm^3$ |
| $W$ | Device Width | 20 nm |
| $t_{Si}$ | Channel thickness | 10 nm |
| $t_{high-k}$ | High k oxide thickness ($HfO_2$) | 1 nm |
| $t_{low-k}$ | Low k oxide thickness ($SiO_2$) | 1 nm |
| $\epsilon_{Si}$ | Dielectric constant of n-silicon | $11.7\epsilon_0$ |
| $\epsilon_{High-k}$ | Dielectric constant of $HfO_2 (High-k)$ | $21\epsilon_0$ |
| $\epsilon_{Low-k}$ | Dielectric constant of $SiO_2 (Low-k)$ | $3.9\epsilon_0$ |
| $n_i$ | Intrinsic carrier concentration | $1.45 \times 10^{10}/cm^3$ |

By this type of gate engineering, a three-step surface potential will be constructed which will supress typical SCEs. This illustrates that SCE parameters like, threshold voltage, DIBL, mobility degradation and off currents will improve overwhelmingly with respect to another JL devices. The impact of lowest work function gate metal at the drain end, reduces electric field and thereby reduces impact ionization, which follows reduction in hot carrier effects (SCEs).

A hetero-dielectric gate oxide stack, a combination of High-k oxide ($HfO_2$) and Low-k oxide ($SiO_2$) to separate gate electrode and channel of the JL MOSFET is also employed in the proposed device. The combination of High-k and Low-k dielectric reduces the fringing effect at the edges, effect of phonon scattering and hence reduces gate leakage current and carrier mobility degradation. This implementation also helps to improve subthreshold behaviour *of the device in nanometre scale.*

Overall, three types of modification are being carried out to construct the proposed novel structure (gate engineering + gate-oxide engineering + channel engineering). Improved characteristics of the proposed novel structure are illustrated in next two sections. Some typical device parameters are specified in Table 1 on basis of which the analytical model is validated. It is clarified that the proposed model is not compared with any experimental results as no experiment has been carried out on the structure.

## 3. Analytical Modelling of Proposed Device

The operating principle of the proposed structure follows basic device physics. Behaviour of this short channel device is analysed under volume conduction mode i.e. the majority carriers of the channel region unanimously flow in the volume (bulk) of the silicon film. For simplicity, it is assumed that, while the device is operating in subthreshold region, the channel is fully depleted and the inversion charge is not considered in the case.

The 2-D channel potential $\varphi_r(x.y)$ can be obtained by solving 2-D Poisson's equations

$$\frac{\partial^2 \varphi_r(x,y)}{\partial x^2} + \frac{\partial^2 \varphi_r(x,y)}{\partial y^2} = \frac{-qN_r}{\epsilon_{Si}} \tag{1}$$

Where subscript r (r=1,2,3), denotes different channel regions, $\varphi_r(x.y)$ are channel potentials and $N_r$ are donor doping densities in different channel regions, q is the electronic charge and $\epsilon_{Si}$ is relative permittivity of silicon. In the proposed device the effective gate oxide stack thickness can be calculated as

$$t_{st} = t_{SiO_2} + t_{SiO_2} \frac{\epsilon_{SiO_2}}{\epsilon_{HfO_2}} \tag{2}$$

This gives the definition of gate oxide stack capacitance, $C_{st} = \frac{\epsilon_{SiO_2}}{t_{st}} \tag{3}$

Now (1) will be solved in different channel regions under gate materials ($M_1$, $M_2$ and $M_3$). The potential distribution $\varphi_r(x.y)$ of graded channel configuration will be approximated as simple parabolic function following Young's approximation [20] (i.e. distribution of electrostatic potential has x dependence only). By means of this approximation, the 2-D Poisson's equation will be reduced to 1-D Poisson's equation and the potential function will be

$$\varphi_r(x,y) = \xi_{r0}(x) + y\xi_{r1}(x) + y^2 \xi_{r2}(x) \tag{4}$$

can be solved by using appropriate boundary conditions.

1. (a) At gate oxide interface, the central potentials $\varphi_c(x)$ under different materials are $\varphi_{cr}(x,0) = \varphi_{cr}(x); r = 1,2,3$ (5)
(b) Similarly, at gate oxide interface, the surface potentials $\varphi_s(x)$ are
$$\varphi_{sr}(x, \tfrac{t_{Si}}{2}) = \varphi_{sr}(x); r = 1,2,3 \tag{6}$$

2. (a) At front gate oxide stack interface, the electric field is continuous in different gate regions $\frac{\partial \varphi_r(x,y)}{\partial y}\Big|_{y=\frac{t_{Si}}{2}} = \frac{C_{st}}{\epsilon_{Si}}\{V'_{gsr} - \varphi_r(x)\}$ (7)

Where r=1,2,3 for $M_1$, $M_2$ and $M_3$ respectively.
(b) At back gate oxide stack interface,
$$\frac{\partial \varphi_r(x,y)}{\partial y}\Big|_{y=-\frac{t_{Si}}{2}} = \frac{-C_{st}}{\epsilon_{Si}}\{V'_{gsr} - \varphi_r(x)\} \tag{8}$$

In (5)-(8),
$$V'_{gsr} = V_{gs} - V_{fb,lr}; r = 1,2,3 \tag{9}$$

Where $V_{gs}$ is gate to source voltage and $V_{fb,lr}$ is flat band voltage under different metal gates. The flat band voltage for n-channel device can be written as

$$V_{fb,lr} = \varphi_{Mr} - \chi_{Si} - \frac{E_g}{2q} + V_t \ln \frac{N_r}{n_i} \; ; r = 1,2,3 \tag{10}$$

Where $\chi_{Si}$ is the electron affinity of silicon, $\frac{E_g}{2q}$ is silicon band gap at room temperature, $n_i$ is intrinsic carrier concentration and $N_r$ is doping concentration in different channel regions, $N_1 < N_2 < N_3$.

(c) The channel potential is symmetrical along y-direction, the electric field must be zero at y=0, i.e., $\frac{\partial \varphi_r(x,y)}{\partial y}|_{y=0} = 0$ \hfill (11)

3. Potential is continuous within channel region, therefore,
   (a) At the interface of two different metal gates, central potentials are continuous i.e.
$$\varphi_{c1}(L_1, 0) = \varphi_{c2}(L_1, 0) \tag{12}$$
and $\varphi_{c2}(L_1+L_2, 0) = \varphi_{c3}(L_1 + L_2, 0)$ \hfill (13)
   (b) Similarly, at the two different metal gate interfaces, surface potentials are also continuous i.e., $\varphi_{s1}(L_1, 0) = \varphi_{s2}(L_1, 0)$ \hfill (14)
and $\varphi_{s2}(L_1+L_2, 0) = \varphi_{s3}(L_1 + L_2, 0)$ \hfill (15)

4. The electric fluxes are also continuous at different metallic interfaces.
   (a) In central region, $\frac{\partial \varphi_{c1}(x)}{\partial x}|_{x=L_1} = \frac{\partial \varphi_{c2}(x)}{\partial x}|_{x=L_1}$ \hfill (16)
and $\frac{\partial \varphi_{c2}(x)}{\partial x}|_{x=L_1+L_2} = \frac{\partial \varphi_{c3}(x)}{\partial x}|_{x=L_1+L_2}$ \hfill (17)
   (b) At surface, $\frac{\partial \varphi_{s1}(x)}{\partial x}|_{x=L_1} = \frac{\partial \varphi_{s2}(x)}{\partial x}|_{x=L_1}$ \hfill (18)
and $\frac{\partial \varphi_{s2}(x)}{\partial x}|_{x=L_1+L_2} = \frac{\partial \varphi_{s3}(x)}{\partial x}|_{x=L_1+L_2}$ \hfill (19)

5. The potential at the source terminal
   (a) $\varphi_{c1}(x = 0) = 0$ \hfill (20)
and (b) $\varphi_{s1}(x = 0) = 0$ \hfill (21)

6. At the drain end, the potential is,
   (a) $\varphi_{c3}(x = L, 0) = V_{DS}; L = L_1 + L_2 + L_3$ \hfill (22)
and (b) $\varphi_{s3}(x = L, 0) = V_{DS}; L = L_1 + L_2 + L_3$ \hfill (23)
where $V_{DS}$ is the applied drain to source bias.

## 3A. Central Potential Modelling

Using boundary condition 1 (a) in (4) gives,
$$\varphi_{cr}(x, 0) = \xi_{r0}(x) = \varphi_{cr}(x) \tag{24}$$
Substituting (7) and (11) into (4), yields,
$$\xi_{r2}(x) = 0 \tag{25}$$
and $\xi_{r3}(x) = \frac{C_{st}}{t_{Si}\epsilon_{Si}} \{V'_{gsr} - \varphi_r(x)\}$ \hfill (26)
Now (4) implies,
$$\varphi_{cr}(x, 0) = \varphi_{cr}(x) + y^2\{V'_{gsr} - \varphi_r(x)\} \tag{27}$$
Differentiating (27) twice with respect to x and y accordingly and putting those values in (1) yields,
$$\frac{\partial^2 \varphi_{cr}(x)}{\partial x^2} - \frac{1}{\lambda_c^2}\{\varphi_{cr}(x) - \sigma_{cr}\} = 0 \tag{28}$$
Where $\lambda_c$ is the scaling length and can be written as-
$$\frac{1}{\lambda_c^2} = \frac{8C_{st}}{4\epsilon_{Si}t_{Si} + t_{Si}^2 C_{st}} \tag{29}$$
and $\sigma_{cr} = V_{gs} - \omega_{cr}; \omega_{cr} = V_{fbr} + \frac{qN_r\lambda_c^2}{\epsilon_{Si}}$ \hfill (30)

The general solution of (28) takes the form,

$$\varphi_{cr}(x) = \alpha_{cr}e^{\frac{x}{\lambda_c}} + \beta_{cr}e^{\frac{-x}{\lambda_c}} + \sigma_{cr} \tag{31}$$

The potential at the centre of the silicon film for different channel regions under different gate metals can be written as-

$$\varphi_{c1}(x) = \alpha_{c1}e^{\frac{x}{\lambda_c}} + \beta_{c1}e^{\frac{-x}{\lambda_c}} + \sigma_{c1}\,; 0 < x < L_1 \text{ under } M_1 \tag{32}$$

$$\varphi_{c2}(x) = \alpha_{c2}e^{\frac{(x-L_1)}{\lambda_c}} + \beta_{c2}e^{\frac{-(x-L_1)}{\lambda_c}} + \sigma_{c2}\,; L_1 < x < L_1 + L_2 \text{ under } M_2 \tag{33}$$

$$\varphi_{c3}(x) = \alpha_{c3}e^{\frac{(x-L_1-L_2)}{\lambda_c}} + \beta_{c3}e^{\frac{-(x-L_1-L_2)}{\lambda_c}} + \sigma_{c3}\,; L_1 + L_2 < x < L_1 + L_2 + L_3 \text{ under } M_3 \tag{34}$$

The co-efficient of $\alpha_{cr}$ and $\beta_{cr}$ have been determined by applying boundary conditions 3-6. Finally, it can be estimated as,

$$\alpha_{c1} = V_{DS} - \sigma_{c3} + \sigma_{c1}e^{\left(\frac{-L}{\lambda_c}\right)} - \frac{(\sigma_{c1}-\sigma_{c2})}{2}\left\{e^{\left(\frac{L_1+L_2}{\lambda_c}\right)} + e^{-\left(\frac{L_1+L_2}{\lambda_c}\right)}\right\} - \frac{(\sigma_{c2}-\sigma_{c3})}{2}\left\{e^{\left(\frac{L_3}{\lambda_c}\right)} + e^{-\left(\frac{L_3}{\lambda_c}\right)}\right\} \tag{35}$$

$$\beta_{c1} = -\sigma_1 - \alpha_1 \tag{36}$$

$$\alpha_{c2} = \alpha_{c1}e^{\left(\frac{L_1}{\lambda_c}\right)} + \frac{\sigma_{c1}-\sigma_{c2}}{2} \tag{37}$$

$$\beta_{c2} = \beta_{c1}e^{\left(\frac{-L_1}{\lambda_c}\right)} + \frac{\sigma_{c1}-\sigma_{c2}}{2} \tag{38}$$

$$\alpha_{c3} = \alpha_{c2}e^{\left(\frac{L_1}{\lambda_c}\right)} + \frac{\sigma_{c2}-\sigma_{c3}}{2} \tag{39}$$

$$\beta_{c3} = \beta_{c2}e^{\left(\frac{-L_2}{\lambda_c}\right)} + \frac{\sigma_{c2}-\sigma_{c3}}{2} \tag{40}$$

## 3B. Surface Potential Modelling

Using boundary condition 1 (b) in (4) gives,

$$\varphi_{sr}\left(x,\frac{t_{Si}}{2}\right) = \xi_{r0}(x) = \varphi_{sr}(x) \tag{41}$$

Substituting (7) and (8) into (4), yields,

$$\xi_{r2}(x) = \frac{C_{st}}{\epsilon_{Si}}\{V'_{gsr} - \varphi_r(x)\} \tag{42}$$

and $\xi_{r3}(x) = \frac{-C_{st}}{t_{Si}\epsilon_{Si}}\{V'_{gsr} - \varphi_r(x)\}$ \hfill (43)

Using (41) – (43) in (4), then differentiating the expression with respect to x and y accordingly and putting those values in (1) gives,

$$\frac{\partial^2 \varphi_{sr}(x)}{\partial x^2} - \frac{1}{\lambda_s^2}\{\varphi_{sr}(x) - \sigma_{sr}\} = 0 \tag{44}$$

Where $\lambda_s$ is the scaling length and can be written as -

$$\frac{1}{\lambda_s^2} = \frac{4C_{st}}{\epsilon_{Si}t_{Si}} \tag{45}$$

and $\sigma_{sr} = V_{gs} - \omega_{sr}; \omega_{sr} = V_{fbr} + \frac{qN_r\lambda_s^2}{\epsilon_{Si}}$ \hfill (46)

The general solution of (28) takes the form,

$$\varphi_{sr}(x) = \alpha_{sr}e^{\frac{x}{\lambda_s}} + \beta_{sr}e^{\frac{-x}{\lambda_s}} + \sigma_{sr} \tag{47}$$

In this case, the expressions for surface potential of the silicon film for different channel regions under different gate metals are given as-

$$\varphi_{s1}(x) = \alpha_{s1}e^{\frac{x}{\lambda_s}} + \beta_{s1}e^{\frac{-x}{\lambda_s}} + \sigma_{s1}\,; 0 < x < L_1 \text{ under } M_1 \tag{48}$$

$$\varphi_{s2}(x) = \alpha_{s2}e^{\frac{(x-L_1)}{\lambda_s}} + \beta_{s2}e^{\frac{-(x-L_1)}{\lambda_s}} + \sigma_{s2}\,; L_1 < x < L_1 + L_2 \text{ under } M_2 \tag{49}$$

$$\varphi_{s3}(x) = \alpha_{s3}e^{\frac{(x-L_1-L_2)}{\lambda_s}} + \beta_{s3}e^{\frac{-(x-L_1-L_2)}{\lambda_s}} + \sigma_{s3}\,; L_1 + L_2 < x < L_1 + L_2 + L_3 \text{ under } M_3 \tag{50}$$

It is noticed that under different gate materials, the same expressions for potentials are obtained in both cases, but the main difference is that scaling lengths are different for central and surface potentials. The coefficients give same expressions as in case of central potential but scaling length is different, that's why value of each coefficient is different from the coefficients obtained in case of central potentials.

### 3C. Central Electric Field Modelling

The expression for central electric field under different gate metals can be obtained by differentiating (32) – (34) as-

$$E_{c1} = \frac{-\alpha_{c1}}{\lambda_c} e^{(\frac{x}{\lambda_c})} + \frac{\beta_{c1}}{\lambda_c} e^{(\frac{-x}{\lambda_c})} \text{ for } 0 < x < L_1 \text{ under } M_1 \quad (51)$$

$$E_{c2} = \frac{-\alpha_{c2}}{\lambda_c} e^{(\frac{x-L_1}{\lambda_c})} + \frac{\beta_{c2}}{\lambda_c} e^{(\frac{-(x-L_1)}{\lambda_c})} \text{ for } L_1 < x < L_1 + L_2 \text{ under } M_2 \quad (52)$$

$$E_{c3} = \frac{-\alpha_{c3}}{\lambda_c} e^{(\frac{x-L_1-L_2}{\lambda_c})} + \frac{\beta_{c2}}{\lambda_c} e^{(\frac{-(x-L_1-L_2)}{\lambda_c})} ; for L_1 + L_2 < x < L_1 + L_2 + L_3$$
$$\text{under } M_3 \quad (53)$$

### 3D. Surface Electric Field Modelling

Similarly surface electric field under different gate metals can be obtained by differentiating (48) – (50) as-

$$E_{s1} = \frac{-\alpha_{s1}}{\lambda_s} e^{(\frac{x}{\lambda_s})} + \frac{\beta_{s1}}{\lambda_s} e^{(\frac{-x}{\lambda_s})} \text{ for } 0 < x < L_1 \text{ under } M_1 \quad (54)$$

$$E_{s2} = \frac{-\alpha_{s2}}{\lambda_s} e^{(\frac{x-L_1}{\lambda_s})} + \frac{\beta_{s2}}{\lambda_s} e^{(\frac{-(x-L_1)}{\lambda_s})} \text{ for } L_1 < x < L_1 + L_2 \text{ under } M_2 \quad (55)$$

$$E_{s3} = \frac{-\alpha_{s3}}{\lambda_s} e^{(\frac{x-L_1-L_2}{\lambda_s})} + \frac{\beta_{s2}}{\lambda_s} e^{(\frac{-(x-L_1-L_2)}{\lambda_s})} ; for L_1 + L_2 < x < L_1 + L_2 + L_3$$
$$\text{under } M_3 \quad (56)$$

### 3E. Threshold Voltage Modelling

In the off-state of a junctionless MOSFET, the channel region with higher doping concentration is fully depleted due to positive ionized dopant and holds a large electric field. When gate voltage increases, the subthreshold electric field in the channel region reduces, thereby creating a neutral region at the center of the channel and volume current starts to flow through the channel region. That's why the threshold voltage expression for the proposed device is derived by considering central potential model.

At threshold voltage, the control gate will be turned on with two adjacent screen gates and the full channel conduction will takes place accordingly. In this case the subthreshold behavior of the device will be determined by the position of minimum central potential, The position of the minimum central potential will be determined under high work function gate metal i.e., $M_1$ at $x = x_{min}$

At $x = x_{min}$, the electric field is assumed to be zero. So, $\frac{d\varphi_{c1}}{dx}|_{x=x_{min}} = 0$ (57)

This implies, $x_{min} = \frac{\lambda_c}{2} \ln(\frac{\beta_1}{\alpha_1})$ (58)

Therefore, the values of minimum central potential,

$$\varphi_{c,min} = 2\sqrt{\alpha_1 \beta_1} + \sigma_1 \quad (59A)$$

In a junctionless MOSFET, the threshold voltage is defined as the required gate voltage for which the minimum central potential can be equated as twice the difference between the Fermi potential and the intrinsic Fermi level in the bulk region of the conducting channel. This definition gives, $\varphi_{c,min} = 2\varphi_F$

(59B)

Equating (59A) and (59B) gives, $2\sqrt{\alpha_1\beta_1} + \sigma_1 = 2\varphi_F$ (60)

By substituting $V_{gs} = V_{th}$ in (56), the expression for threshold voltage can be written as,

$$V_{th} = \frac{-Y+\sqrt{(Y^2-4XZ)}}{2X}$$ (61)

Where X, Y and Z are defined in "Appendix".

For long channel device, $L \to \infty$ then $\alpha_1 \sim 0$ and $\beta_1 \sim 0$. In case of a long channel device, the threshold voltage can be expressed as

$$Lim\ V_{th}|_{L\to\infty} = V_{th,L} = 2\varphi_F + \omega_1$$ (62)

### 3F. Threshold Voltage Roll-off

Threshold voltage roll off occurs by virtue of short channel effect of the device and can be defined as the difference between the threshold voltage of a short channel device to that of a long channel device. Threshold voltage is independent of channel length for any long channel MOSFET.

$$V_{th,roll-off} = V_{th} - V_{th,L}$$ (63)

### 3G. Threshold Voltage Modelling

DIBL of a junctionless MOSFET is defined as the ratio of the change in threshold voltage (high to low) to change in drain bias (low to high).

$$DIBL = \frac{(V_{th,high} - V_{th,low})}{(V_{DS,low} - V_{DS,high})}$$ (64)

### 3H. Subthreshold Swing (SS)

This an important parameter to determine switching behavior of the device and can be expressed as,

$$SS = \frac{V_{th} \ln 10}{\frac{\partial \varphi_{c1,min}}{\partial V_{gs}}}$$ (65)

Using (58) in (63),

$$SS = \frac{V_{th} \ln 10 \sqrt{\alpha_1\beta_1}}{\{\alpha_1 \frac{\partial \beta_1}{\partial V_{gs}} + \beta_1 \frac{\partial \beta_1}{\partial V_{gs}} + \sqrt{\alpha_1\beta_1}\}}$$ (66)

### 3I. Drain Current Model

### I. Subthreshold Region

Subthreshold drain current of a junctionless MOSFET is characterized as the current flow between the source and drain at $V_{gs} < V_{th}$ and in this case diffusion current plays an important role. Following continuity equation, the total drain current can be written as [21],

$$J_n(x,y) = -q\mu_{eff}n_r(x,y)\frac{d\varphi_n(x)}{dx}$$ (67)

Where $\mu_{eff}$ is the effective mobility of electron in channel region, $\varphi_n(x)$ is quasi-Fermi potential of electron which is constant in y-direction of the channel and $n_r(x,y)$ is the carrier concentration given as-

$$n_r(x,y) = \frac{n_i^2}{N_r} \exp\left(\frac{\varphi_r(x,y)}{V_{th}}\right)$$ (68)

Integrating $J_n(x,y)$, the subthreshold current can be written as,

$$I_{DS\,sub}(x) = -W \int_{\frac{t_{Si}}{2}}^{\frac{-t_{Si}}{2}} J_n(x,y) dy \tag{69}$$

Now, by following Pao-Sah's double integral [22], the expression for subthreshold current can be formulated as,

$$I_{DS\,sub} = \frac{W\mu_{eff} k_B T[1-e^{\frac{-V_{DS}}{V_{th}}}]}{\{\int_0^{L_1} Q_1^{-1}(x)dx + \int_{L_1}^{L_1+L_2} Q_2^{-1}(x)dx + \int_{L_1+L_2}^{L_1+L_2+L_3} Q_3^{-1}(x)dx\}} \tag{70}$$

Where the depletion charge $Q_j(x)$ is given as,

$$Q_j(x) = -q \int_{\frac{-t_{Si}}{2}}^{\frac{t_{Si}}{2}} \frac{n_i^2}{N_r} \exp\left[\frac{q}{k_B T}\{\varphi(x,y) - \varphi_n(x)\}dy\right] \approx \frac{t_{Si} q n_i^2 e^{\frac{\varphi_{cr,min}}{V_{th}}}}{2N_r} \tag{71}$$

(Subthreshold current generally arises at, $x = x_{min}$)

Solving (64) with (65), the expression for subthreshold current is obtained,

$$I_{DS\,sub} = \frac{W\mu_{eff} k_B T n_i^2 [1-e^{\frac{-V_{DS}}{V_{th}}}]}{\{\frac{N_1 L_1}{e^{\frac{\varphi_{c1,min}}{V_{th}}}} + \frac{N_2 L_2}{e^{\frac{\varphi_{c2,min}}{V_{th}}}} + \frac{N_3 L_3}{e^{\frac{\varphi_{c3,min}}{V_{th}}}}\}} \tag{72}$$

Barsan et al. [23] shows that a dual gate MOSFET in subthreshold region create a potential barrier or potential well under second gate and the sub threshold current is either controlled by potential barrier or drained by potential well. Following the same principal, the sub threshold current of our proposed device will be controlled by the highest potential barrier as $\varphi_{c3,min} > \varphi_{c2,min} > \varphi_{c1,min}$. The expression for subthreshold current can be written as

$$I_{DS\,sub} = \frac{W\mu_{eff} k_B T n_i^2 [1-e^{\frac{-V_{DS}}{V_{th}}}]}{N_1 L_1} e^{\frac{\varphi_{c1,min}}{V_{th}}} \tag{73}$$

## II. Linear Region

The proposed graded channel device can be treated as three sub-devices connected in series and each segment has different threshold voltages. To construct a compact model, we have to obtain individual drain current for each segment. In linear region, drift current is most effective and can be written as [24, 25]

$$I_{DS}(x) = \frac{W Q_{nr} \mu_{nr} \frac{d\varphi(x)}{dx}}{\{1 + \frac{1}{E_{sat}}[\frac{d\varphi(x)}{dx}]\}} \tag{74}$$

Where W is the width of the device, $\varphi(x)$ is channel potential, $\frac{d\varphi(x)}{dx}$ is channel electric field, $E_{sat,r}(= \frac{2v_{sat}}{\mu_{nr}})$ is the critical electric field and $v_{sat}$ is the saturation velocity of electron.

The surface charge density

$$Q_{nr}(x) = C_{st}[V_{gs} - V_{th,r} - \varphi_r(x)] \tag{75}$$

The mobility in the channel region is given by

$$\mu_{nr} = \frac{\mu_i}{1 + \theta_r(V_{gs} - V_{th,r})} \tag{76}$$

Where $\theta_r$ is the fitting parameters and $\mu_i$ is mobility of impurity density and can be expressed as-

$$\mu_i = \frac{\mu_n}{\sqrt{1 + \frac{N_r}{\{N_{ref} + \frac{N_r}{s}\}}}} \tag{77}$$

Where $\mu_n(= 670 cm^2/Vsec)$ is mobility of electron in n-channel and $s(= 350)$ is the fitting parameter for trade-off between impurity scattering and phonon, $N_r$ is the doping concentration in different channel segment and $N_{ref}(= 1.072 \times 10^{17} cm^{-3})$.

For the new triple metal gated device (solving (72)),

$$\frac{\mu_1(V_{gs} - V_{th1})V_1}{L_1} = \frac{\mu_2(V_{gs} - V_{th2})V_2}{L_2} = \frac{\mu_3(V_{gs} - V_{th3})V_3}{L_3} \tag{78}$$

Where $V_1, V_2$ and $V_3$ are voltage drop in three regions and $V_1 + V_2 + V_3 = V_{DS}$

Substituting (70), (71) and (72) in (68) and integrating, the expression for drain current in linear region can be written as

$$I_{DS} = \frac{C_{st} W \mu_{eff} (V_{gs} - V_{th} - 0.5 V_{DS}) V_{DS}}{L(1 + \frac{V_{DS}}{LE_{eff}})} \quad (79)$$

where $\mu_{eff} = \frac{L}{\{\frac{L_1}{\mu_1} + \frac{L_2}{\mu_2} + \frac{L_3}{\mu_3}\}}$ (80)

and $E_{eff} = \frac{2 v_{sat}}{\mu_{eff}}$ (81)

## III. Saturation Region

In the saturation region, the expression for drain current is

$$I_{d\,sat} = W V_{sat} Q_{sat} \quad (82)$$

where $Q_{sat}$ is saturation current density and can be written as

$$Q_{sat} = C_{st}(V_{gs} - V_{th} - V_{D\,sat}) \text{ at } V_{DS} = V_{D\,sat} \quad (83)$$

where drain saturation voltage $V_{D\,sat} = \frac{(V_{gs} - V_{th})}{\{1 + \frac{(V_{gs} - V_{th})}{LE_{eff}}\}}$ (84)

## 4. Results and Discussion

The proposed device performance is validated and analyzed by simulation using Wolfram Mathematica software with suitable conditions. To validate this novel device some optimized values of basic device parameters are used. The specialty of this device is the total gate length is reduced below 30nm with the consideration of gate length in the ratio $L_1 : L_2 : L_3 = 1 : 1 : 1$. If the gate length ($L_1$) of the control gate under metal $M_1$ is reduced below 9 nm, the position of minimum central potential as well as minimum surface potential will be shifted towards the source which will results intense DIBL (drain induced barrier lowering). On the other hand, the reduction of gate length in the portion of second screen gate ($L_3$) in drain end, will result an exacerbate hot carrier effect. Considering the basic physics of metal semi-conductor interface (the elaborate description is omitted in this paper), optimized values of gate work functions are chosen and declared in Table 1. It is also noted that in the proposed device, the measured scaling length for central potential model is greater than that of surface potential model, i.e., $\lambda_c > \lambda_s$. This indicates that the device utilizes more control in drain side over the channel than the gate control. This section contains analysis of different graphical plots which illustrate various characteristic parameters involving SCEs of the proposed device.

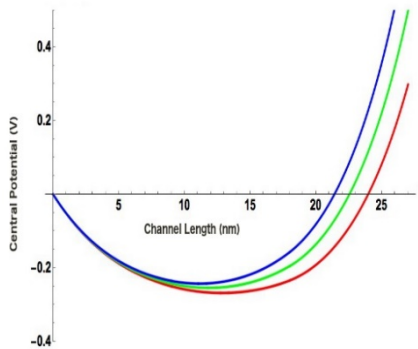
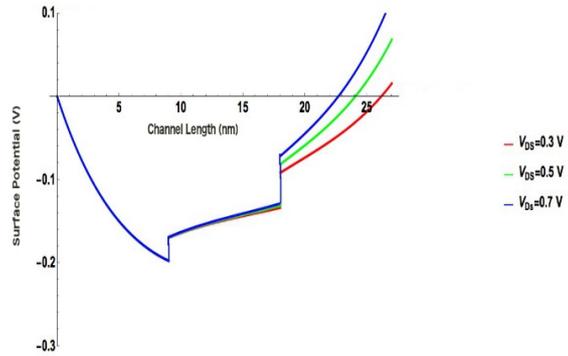

**Fig 2: Variation of Central Potential with channel (gate) length for different drain bias**

**Fig 3: Variation of Surface Potential with channel (gate) length for different drain bias**

Fig 2 shows the central potential profile of the proposed device along channel length (gate length) for different drain to source voltages ($V_{DS}$) with a constant gate to source bias ($V_{gs}$). From the plot, it is clearly observed that the position of minimum central potential incline towards source side for this graded channel device. This type of potential profile is also helpful to minimize the most hazardous SCE, like DIBL and it also reduces threshold voltage roll off. Another point also noticed from this plot is, for lower drain voltage, the position of minimum potential occurs under gate $M_1$, the metal gate with higher work function value (control gate), which is an important prior assumption for formulation of threshold voltage expression of the device.

Fig 3 demonstrates the surface potential profile along different position of the entire channel length with variation of drain voltage ($V_{DS}$) for a fixed gate to source voltage ($V_{GS}$). This potential profile indicates step change at two different metallic interfaces with high-low work functions and with different doping concentration in the channel region of this graded channel device. Step change of potential profile implies carrier velocity enhancement which indicates carrier transport efficiency. Another crucial feature of this plot is, the minimum surface potential occurs under highest work function metal gate $M_1$(control gate).

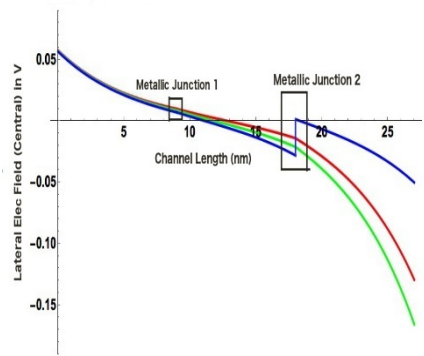
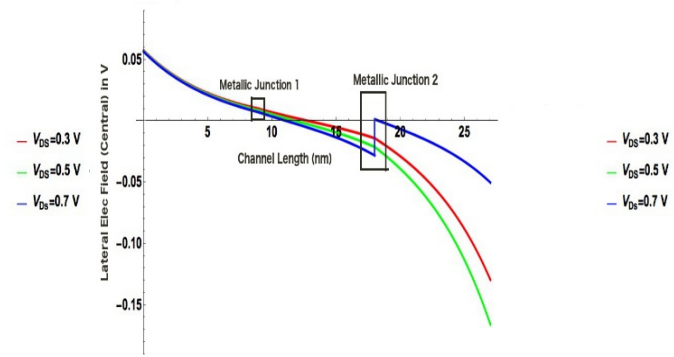

**Fig 4.** Variation of Lateral Electric Field (in Central region) with channel (gate) length for different drain bias

**Fig. 5:** Variation of Lateral Electric Field (in Surface region) with channel (gate) length for different drain bias

Fig 4 illustrates variation of lateral electric field for central potential as a function of different channel position with variation of drain voltage ($V_{DS}$) for a fixed gate voltage ($V_{GS}$)t is clearly seen that, two peaks are formed at the interface of different metal gates. The most interesting point of the plot is that, the profile exhibits low electric field at the drain end, which indicates reduction of hot carrier effect (HCE) at the drain end. Introduction of graded channel with hetero-dielectric gate oxide stack in triple metal double gate junctionless MOSFET boosts hot carrier immunity and it also enhance device reliability in low power application.

Fig 5 depicts lateral electric field profile for surface potential at different channel positions with varying drain voltage ($V_{DS}$) for a fixed gate voltage ($V_{GS}$). This plot is similar in nature as in Fig 4. Surface electric filed profile also indicates hot carried reduction in the drain side of our proposed device.

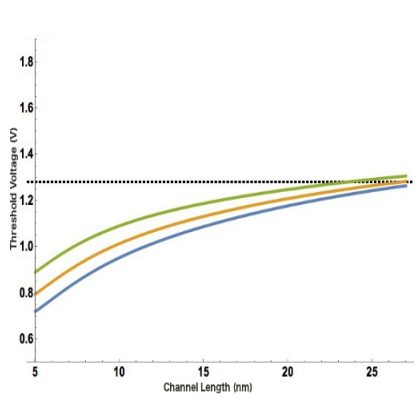 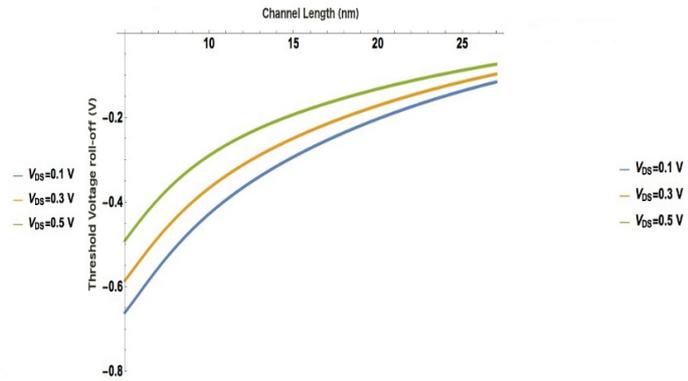

**Fig 6. Variation of Threshold voltage ($V_{th}$) with channel position for different drain different drain bias**

**Fig. 7: Variation of Threshold voltage ($V_{th}$) roll-off with different channel (gate) length for bias**

For a junctionless MOSFET, at the centre of the channel region a leakage path is created and hence central potential based threshold voltage model has been formulated for the proposed junctionless structure. To ameliorate threshold voltage, a thin layer of gate oxide is preferred. The proposed device is consisting of a combination of a high-k oxide layer and a low-k oxide layer, both with 1nm thickness and this combination results an equivalent gate oxide stack thickness of approximately 1 nm which is thinner than gate oxide stack thickness proposed by P wang et al. [9] in case of dual metal double gate JL-MOSFET. Fig 6, indicates the variation of threshold voltage along different position of the channel length for variable drain voltage ($V_{DS}$) with a fixed gate voltage ($V_{GS}$). The plot shows slightly high threshold voltage for the proposed device. It can be argued qualitatively that, the threshold voltage of the proposed device is quite better in comparison with DMDG junctionless MOSFET proposed by P Wang et al. [9] and GC-DMDG JL MOSFET proposed by Pathak et al.[14].This can be a remarkable advantage as high threshold voltage indicates reduction of leakage current which is a basic requirement for low power application. The threshold voltage has been calculated considering potential under highest work function metal gate, but the plot indicates that there is a little variation of threshold voltage for different drain bias. This actually means, there is both gate control as well as drain control over channel region. If there is no gate control over channel region, threshold voltage could be decreased for variable drain bias.

Fig 7 represents threshold voltage roll off depends on channel length for variation of drain bias. Decreasing channel length implies increase in threshold voltage roll off. This is a typical SCE in short channel device. Reduction of threshold voltage roll off indicates better control over SCEs and enhancement of performance of this new device in comparison with JL DMDG stack MOSFET proposed by S Darwin and Arun Samuel TS et al. [10] and DMDG JL MOSFET was proposed by P Wang et al. [9].

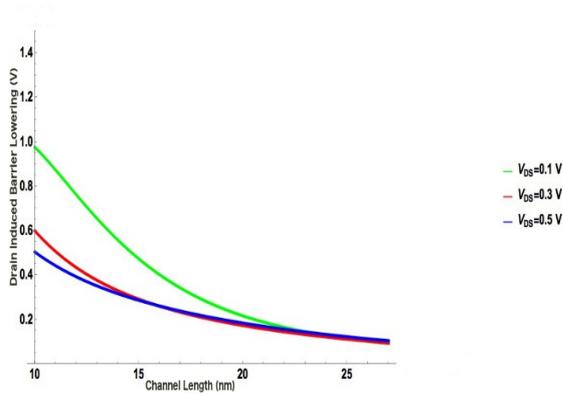
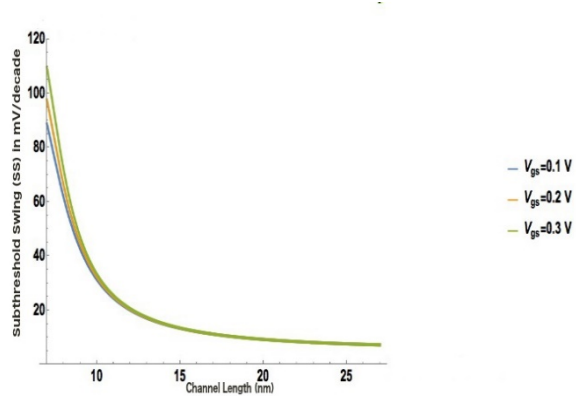

**Fig 8. Variation of Drain Induced Barrier Lowering (DIBL) with respect to different channel position for different drain bias**

**Fig. 9: Variation of Subthreshold Swing ($SS$) withrespect to different channel position for differentdrain bias**

Fig 8 represents Drain-Induced-Barrier-Lowering as a function of channel length for variable drain bias. The plot reflects screening effect to suppress the drain voltage into the channel region and results the lowering of DIBL. Dual step nature of potential profile due to different work function metal gate with graded channel combination helps us to reduce DIBL effect of the proposed device and approaches minimum position of potential towards the source end.

Fig 9 shows the variation of subthreshold swing with channel length for different drain bias with a fixed low gate voltage. The plots clearly indicates that, the subthreshold swing for our propose device is below 120mV/decade. Basically, Subthreshold swing depicts the switching characteristics of the device. This type of SS value indicates not less, but moderate off state leakage current. Moreover, in comparison between this channel engineered novel device and TMDG JL MOSFET suggested by Darwin S and Arun Samuel et al. [12], the novel device reveals better subthreshold swing (SS value) which indicates better suppression of hazardous SCEs in nanometer regime.

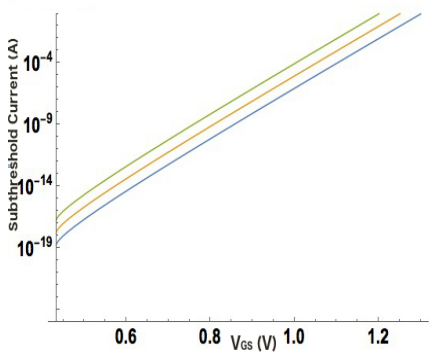
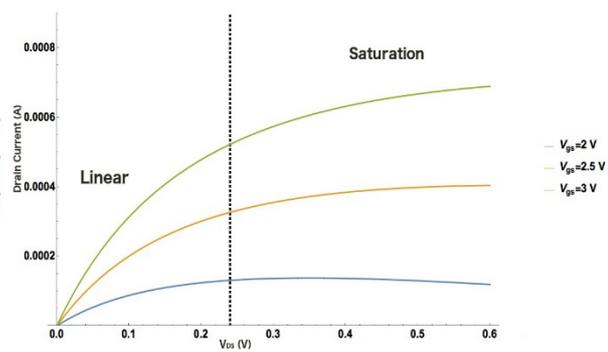

**Fig 10. Variation of Subthreshold current with respect to gate to source voltage ($V_{gs}$) different for different drain bias**

**Fig 11: Variation of drain current with respect to drain to source voltage ($V_{DS}$) with gate to source voltage ($V_{gs}$)**

Fig 10 illustrates the behavior of subthreshold current of the proposed device with variation of gate to source voltage from zero ($V_{DS}$) to threshold voltage ($V_{DS} = V_{th}$), for different drain bias. From this plot, most important figure of merit of the device $I_{on}$ / $I_{off}$ ratio can be determined. Higher ON-current always results better performance in ultra-low power application which indicates better ON-current characteristics of this novel device in comparison with GC-DMDG JLT constructed by Pathak et al [14] as well as this novel device exhibits better ON-OFF ratio than TMDG JL MOSFET predicted by Darwin S and Arun Samuel et al. [12]. It is to be pointed out that the plot of subthreshold current with increasing gate voltage ensures excellent reduction in leakage current and very high $I_{on}/I_{off}$ ratio. So, the device has better SCE immunity and more gate control in sub threshold region. The device has good switching characteristic which is a crucial requirement in low power application. Subthreshold current increases with increase in drain current. The inference given for subthreshold current behavior of the device is that incorporation of graded channel doping profile and thin hetero-dielectric gate oxide are the main key points to improve subthreshold current profile of the proposed device.

Fig 11 demonstrate the impact of drain bias on drain current of the proposed device in linear and saturation regions. Increase in $V_{GS}$ results increase in drain current with drain to source voltage. With increase in drain bias, drain current is increasing in linear region, and after reaching the maximum value the drain current saturates at $V_{DS} > V_{DS,Sat}$. Qualitative comparison of the parameter, DIBL of this new device with GC-DMDG JLT constructed by Pathak et al [14], DMDG JL MOSFET proposed by P Wang et al. [9] and TMDG JL MOSFET suggested by Darwin S and Arun Samuel et al. [12], indicates improved response of this crucial parameter in this proposed device. Less DIBL and better immunities against other SCEs and HCEs of the device results enhancement of drain current.

## 4. Conclusion

A theoretical model of triple metal double gate graded channel junctionless MOSFET with hetero dielectric gate oxide stack below 30 nm regime has been studied in this work. The most relevant advantage of this newly proposed device is that it is more advanced (in respect of sub-threshold current, threshold voltage and other improved SCEs) and reliable junctionless device than any other junctionless MOSFETs available in literature for ultra-low power applications in neoteric world of microelectronics and nanotechnology. But the basic characteristics of this device are presumed upon considering some optimized values of device parameters which are the limitation of this novel device. Moreover, in this analytical model, Quantum Mechanical Effect (QEM) is ignored in overall derivation since the channel (gate) length has been considered greater than 25nm and silicon channel is 10 nm. Both central and surface potential profile of this fully depleted n-channel device have been derived considering Young's approximation by virtue of which the potential function behaves as simple parabolic function. Use of triple metal double gate in junctionless MOSFET structure yields step potential profile on the surface of the device. Reduced electric field in the drain end is the most significant outcome of incorporation of graded channel which clearly indicates reduction in HCEs in the drain region. The overall analysis of device performance demonstrates improvement in SCEs like DIBL and threshold voltage roll off. Influence of hetero dielectric gate oxide stack has been seen in subthreshold drain current characteristics. Analysis of Subthreshold Swing (SS) indicates better switching performance. Derivation of comprehensive subthreshold current model results reduction in leakage current as well as higher ratio. Subthreshold behavior of the device over wide range of voltage may be helpful in future development of ultra-low power Nano scale devices below 30 nm regime with inclusion of QEM. Moreover, this simple model can easily be understood and no rigorous mathematical calculation is involved in the entire formulation and analysis.

## Acknowledgment

We acknowledge the use of licensed Wolfram Mathematica Software for the plots. Ankush Ghosh thankfully acknowledges the support received from the Neotia University vide File no. R&D/2020/F8 Dated 01.04.2020.

## Appendix

$$X = \left\{\exp\left(\frac{L}{\lambda_c}\right) + \exp\left(\frac{-L}{\lambda_c}\right) - 2\right\} - \left\{0.25\left[\exp\left(\frac{L}{\lambda_c}\right) + \exp\left(\frac{-L}{\lambda_c}\right)\right]^2\right\} \tag{A1}$$

$$Y = J_1 + J_2 + J_3 + J_4 + J_5 \tag{A2}$$

$$J_1 = \left\{-V_{DS} - \omega_1 - \omega_3 - \frac{1}{2}(\omega_1 - \omega_2)\left[\exp\left(\frac{L_1+L_2}{\lambda_c}\right) + \exp\left(-\frac{L_1+L_2}{\lambda_c}\right)\right] - \frac{1}{2}(\omega_2 - \omega_3)\left[\exp\left(\frac{L_3}{\lambda_c}\right) + \exp\left(-\frac{L_3}{\lambda_c}\right)\right]\right\}\left\{\exp\left(\frac{L}{\lambda_c}\right) + \exp\left(-\frac{L}{\lambda_c}\right)\right\} \quad (A3)$$

$$J_2 = (\omega_1 - \omega_2)\left[\exp\left(\frac{L_1+L_2}{\lambda_c}\right) + \exp\left(-\frac{L_1+L_2}{\lambda_c}\right)\right] \quad (A4)$$

$$J_3 = (\omega_2 - \omega_3)\left[\exp\left(\frac{L_3}{\lambda_c}\right) + \exp\left(-\frac{L_3}{\lambda_c}\right)\right] \quad (A5)$$

$$J_4 = 2V_{DS} + 2\omega_1 + 2\omega_3 \quad (A6)$$

$$J_5 = \frac{(2\varphi_F + \omega_1)}{2}\left[\exp\left(\frac{L}{\lambda_c}\right) - \exp\left(-\frac{L}{\lambda_c}\right)\right]^2 \quad (A7)$$

$$Z = K_1 + K_2 + K_3 + K_4 + K_5 \quad (A8)$$

$$K_1 = \left\{\omega_1 V_{DS} + \omega_1\omega_3 + \frac{1}{2}\omega_1(\omega_1 - \omega_2)\left[\exp\left(\frac{L_1+L_2}{\lambda_c}\right) + \exp\left(-\frac{L_1+L_2}{\lambda_c}\right)\right] + \frac{1}{2}\omega_1(\omega_2 - \omega_3)\left[\exp\left(\frac{L_3}{\lambda_c}\right) + \exp\left(-\frac{L_3}{\lambda_c}\right)\right]\right\}\left\{\exp\left(\frac{L}{\lambda_c}\right) + \exp\left(-\frac{L}{\lambda_c}\right)\right\} \quad (A9)$$

$$K_2 = \left\{-V_{DS}(\omega_1 - \omega_2) - \omega_3(\omega_1 - \omega_2) - \frac{(\omega_1-\omega_2)^2}{4}\left[\exp\left(\frac{L_1+L_2}{\lambda_c}\right) + \exp\left(-\frac{L_1+L_2}{\lambda_c}\right)\right] - \frac{(\omega_1-\omega_2)(\omega_2-\omega_3)}{2}\left[\exp\left(\frac{L_3}{\lambda_c}\right) + \exp\left(-\frac{L_3}{\lambda_c}\right)\right]\right\}\left[\exp\left(\frac{L_1+L_2}{\lambda_c}\right) + \exp\left(-\frac{L_1+L_2}{\lambda_c}\right)\right] \quad (A10)$$

$$K_3 = \left\{-V_{DS}(\omega_2 - \omega_3) - \omega_3(\omega_2 - \omega_3)\frac{(\omega_1-\omega_2)^2}{4}\left[\exp\left(\frac{L_3}{\lambda_c}\right) + \exp\left(-\frac{L_3}{\lambda_c}\right)\right]\right\}\left[\exp\left(\frac{L_3}{\lambda_c}\right) + \exp\left(-\frac{L_3}{\lambda_c}\right)\right] \quad (A11)$$

$$K_4 = -V_{DS}^2 - 2\omega_3 V_{DS} - \omega_1^2 - \omega_3^2 \quad (A12)$$

$$K_5 = \frac{(2\varphi_F + \omega_1)^2}{4}\left[\exp\left(\frac{L}{\lambda_c}\right) - \exp\left(-\frac{L}{\lambda_c}\right)\right]^2 \quad (A13)$$